\documentclass[preprint]{emulateapj}

\newcommand{\nustar}{{\it NuSTAR}}
\newcommand{\swift}{{\it Swift}}
\newcommand{\chandra}{{\it Chandra}}
\newcommand{\xmm}{{\it XMM-Newton}}

\def\arcsec{$\,^{\prime\prime}$}

\def\fluxu{erg~s$^{-1}$~cm$^{-2}$}

\def\simless{\mathbin{\lower 3pt\hbox
   {$\rlap{\raise 5pt\hbox{$\char'074$}}\mathchar"7218$}}} 
\def\simgreat{\mathbin{\lower 3pt\hbox
   {$\rlap{\raise 5pt\hbox{$\char'076$}}\mathchar"7218$}}} 
\def\cmsq{cm$^{-2}$}

\slugcomment{Accepted by ApJ.}

\shorttitle{Resolving the X-ray Background with {\em NuSTAR}}
\shortauthors{Harrison et al.}

\begin{document}


\title{The NuSTAR Extragalactic Surveys: The Number Counts of Active Galactic Nuclei and the Resolved Fraction of the Cosmic X-ray Background}

\author{F. A. Harrison\altaffilmark{1}}
\author{J. Aird\altaffilmark{2,3}}
\author{F. Civano\altaffilmark{4}}
\author{G. Lansbury\altaffilmark{3}}
\author{J. R. Mullaney\altaffilmark{5}}
\author{D. R. Ballantyne\altaffilmark{6}}
\author{D. M. Alexander\altaffilmark{3}}
\author{D. Stern\altaffilmark{7}}
\author{M. Ajello\altaffilmark{8}}
\author{D. Barret\altaffilmark{9}}
\author{F. E. Bauer\altaffilmark{10,11,12}}
\author{M. Balokovi\'{c}\altaffilmark{1}}
\author{W. N. Brandt\altaffilmark{13,14,15}}
\author{M. Brightman\altaffilmark{1}}
\author{S. E. Boggs\altaffilmark{16}}
\author{F. E. Christensen\altaffilmark{17}}
\author{A. Comastri\altaffilmark{18}}
\author{W. W. Craig\altaffilmark{16,19}}
\author{A. Del Moro\altaffilmark{3}}
\author{K. Forster\altaffilmark{1}} 
\author{P. Gandhi\altaffilmark{20}}
\author{P. Giommi\altaffilmark{21}}
\author{B. W. Grefenstette\altaffilmark{1}}  
\author{C. J. Hailey\altaffilmark{22}} 
\author{R.~C.~Hickox\altaffilmark{23}}
\author{A. Hornstrup\altaffilmark{17}} 
\author{T. Kitaguchi\altaffilmark{24}}
\author{J. Koglin\altaffilmark{26}}
\author{B. Luo\altaffilmark{13,14,15}}
\author{K. K. Madsen\altaffilmark{1}}
\author{P. H. Mao\altaffilmark{1}} 
\author{H. Miyasaka\altaffilmark{1}} 
\author{K. Mori\altaffilmark{22}}
\author{M. Perri\altaffilmark{21,28}} 
\author{M. Pivovaroff\altaffilmark{19}} 
\author{S. Puccetti\altaffilmark{21,28}} 
\author{V. Rana\altaffilmark{1}}
\author{E. Treister\altaffilmark{25}} 
\author{D. Walton\altaffilmark{6}}
\author{N. J. Westergaard\altaffilmark{17}}
\author{D. Wik\altaffilmark{27}}
\author{L. Zappacosta\altaffilmark{28}}
\author{W. W. Zhang\altaffilmark{27}}
\author{A. Zoglauer\altaffilmark{16}} 
\altaffiltext{1}{Cahill Center for Astronomy and Astrophysics, California Institute of Technology, Pasadena, CA 91125}
\altaffiltext{2}{Institute of Astronomy, University of Cambridge, Madingley Road, Cambridge, CB3 0HA, U.K.}
\altaffiltext{3}{Centre of Extragalactic Astronomy, Department of Physics, Durham University, Durham, DH1 3LE, U.K.}
\altaffiltext{4}{Yale Center for Astronomy and Astrophysics, 260 Whitney Avenue, New Haven, CT 06520, USA}
\altaffiltext{5}{The Department of Physics and Astronomy, The University of Sheffield, Hounsfield Road, Sheffield S3 7RH, U.K.}
\altaffiltext{6}{Center for Relativistic Astrophysics, School of Physics, Georgia Institute of Technology, Atlanta, GA 30332, USA}
\altaffiltext{7}{Jet Propulsion Laboratory, California Institute of Technology, 4800 Oak Grove Drive, Pasadena, CA 91109, USA}
\altaffiltext{8}{Department of Physics and Astronomy, Clemson University, Kinard Lab of Physics, Clemson, SC 29634-0978, USA}
\altaffiltext{9}{Universite de Toulouse; UPS-OMP; IRAP; Toulouse, France \& CNRS; Institut de Recherche en Astrophysique et Plan\'etologie; 9 Av. colonel Roche, BP 44346, F-31028 Toulouse cedex 4, France}
\altaffiltext{10}{Instituto de Astrof\'{i}sica, Facultad de F\'{i}sica, Pontificia Universidad Cat\`{o}lica de Chile, 306, Santiago 22, Chile}
\altaffiltext{11}{Millennium Institute of Astrophysics, Santiago, Chile}
\altaffiltext{12}{Space Science Institute, 4750 Walnut Street, Suite 205, Boulder, Colorado 80301, USA}
\altaffiltext{13}{Department of Astronomy and Astrophysics, The Pennsylvania State University, 525 Davey Lab, University Park, PA 16802, USA}
\altaffiltext{14}{Institute for Gravitation and the Cosmos, The Pennsylvania State University, University Park, PA 16802, USA}
\altaffiltext{15}{Department of Physics, The Pennsylvania State University, University Park, PA 16802, USA}
\altaffiltext{16}{Space Sciences Laboratory, University of California, Berkeley, CA 94720, USA}
\altaffiltext{17}{DTU Space - National Space Institute, Technical University of Denmark, Elektrovej 327, 2800 Lyngby, Denmark}
\altaffiltext{18}{INAF Osservatorio Astronomico di Bologna, via Ranzani 1, I-40127, Bologna, Italy}
\altaffiltext{19}{Lawrence Livermore National Laboratory, Livermore, CA 94550, USA}
\altaffiltext{20}{School of Physics \& Astronomy, University of Southampton, Highfield, Southampton SO17 1BJ, U.K.}
\altaffiltext{21}{ASI Science Data Center (ASDC), via del Politecnico, I-00133 Rome, Italy}
\altaffiltext{22}{Columbia Astrophysics Laboratory, Columbia University, New York, NY 10027, USA}
\altaffiltext{23}{Department of Physics and Astronomy, Dartmouth College, 6127 Wilder Laboratory, Hanover, NH 03755, USA}
\altaffiltext{24}{Department of Physical Science, Hiroshima University, 1-3-1 Kagamiyama, Higashi-Hiroshima, Hiroshima 739-8526, Japan}
\altaffiltext{25}{Universidad de Concepci\`{o}n, Departamento de Astronom\'{i}a, Casilla 160-C, Concepci\`{o}n, Chile}
\altaffiltext{26}{Kavli Institute for Particle Astrophysics and Cosmology, SLAC National Accelerator Laboratory, Menlo Park, CA 94025, USA}
\altaffiltext{27}{NASA Goddard Space Flight Center, Greenbelt, MD 20771, USA}
\altaffiltext{28}{INAF-Osservatorio Astronomico di Roma, via di Frascati 33, I-00078, Monte Porzio Catone, Italy}

\begin{abstract}
We present the 3 -- 8~keV and 8 -- 24~keV number counts of active galactic nuclei (AGN) identified in the \nustar\ 
extragalactic surveys.   \nustar\ has now resolved 33 - 39\% of the X-ray background in the 8 -- 24~keV band, directly identifying
AGN with obscuring columns up to $\sim 10^{25}$~\cmsq.    
In the softer 3~--~8~keV band the number counts are in general agreement with those measured by \xmm\ and \chandra\ over the 
flux range $5 \times 10^{-15} \simless S {\rm(3 - 8 keV)} /$\fluxu$\simless 10 ^{-12}$ probed by \nustar.  
In the hard 8 -- 24~keV band \nustar\ probes fluxes over the range $2 \times 10^{-14} \simless S {\rm(8 - 24~keV)}/$\fluxu$ \simless 10 ^{-12}$, a
factor $\sim$100 fainter than previous measurements. 
The 8~--~24~keV number counts match predictions from AGN population-synthesis models, directly confirming the existence of a population of obscured and/or hard X-ray sources inferred from the shape of the integrated cosmic X-ray background. 
The measured \nustar\ counts lie significantly above simple extrapolation with a Euclidian slope to low flux of the \swift/BAT 15 -- 55~keV number counts measured at higher fluxes  ($S (15 - 55~{\rm keV}) \simgreat 10^{-11}$~\fluxu ), reflecting the evolution of the AGN population between the \swift/BAT local  ($z < 0.1$)  sample and \nustar's $z \sim 1$ sample. 
CXB synthesis models, which account for AGN evolution, lie above the \swift/BAT measurements, suggesting that they do not fully capture the evolution of obscured AGN at low redshifts.
\end{abstract}


\section{Introduction}

A complete census of accreting supermassive black holes (SMBH) throughout cosmic time is necessary to quantify the efficiency of accretion, 
which is believed to drive the majority of SMBH growth \citep[e.g.][]{soltan82,yt02,dcs+08,mh08}.   X-ray emission is nearly universal from the luminous Active Galactic Nuclei (AGN) that signal the most rapid SMBH growth phases, making surveys in the X-ray band particularly efficient at identifying accreting
SMBH.    Unlike optical and infrared light, 
X-rays are not diluted by host-galaxy
emission, which is generally weak above $\sim$1~keV.    X-rays are also penetrating, and hard ($\simgreat$ 10~keV)  X-rays are visible through columns up to $N_{\rm H} \sim 10^{25}$~\cmsq. 
For even higher columns AGN can be identified through scattered X-rays, although the emission becomes progressively weaker with increasing column.  

Cosmic X-ray surveys with \chandra\ and \xmm\  have provided measurements of
the demographics of the AGN population and its evolution in the 0.1 -- 10~keV band out to large cosmic distances \citep[see][for a recent review]{ba15}. These surveys are sufficiently complete over a broad enough range
of luminosity and redshift that many fundamental questions regarding AGN evolution can be addressed.
In the deepest fields,  $>80$\% of the 2 -- 10~keV Cosmic X-ray Background (CXB) has been resolved into individual objects \citep{hm06,ba15}.    

At X-ray energies above 10~keV the observational picture is far less complete.    Coded-mask instruments such as {\em INTEGRAL} and {\em Swift}/BAT have probed the demographics of hard X-ray emitting AGN in the very local universe, to redshifts $z\simless 0.1$ \citep{tm+08,bs+09}.
The fraction of the CXB resolved by these instruments at its peak intensity (20 -- 30 keV) is $\sim$1\% \citep{kr+11,aa+12,vm+13}.
Thus, until now,  samples of AGN selected at $> 10$~keV, which are inherently less biased by obscuration than those  at lower energy,
could not be used to probe AGN demographics, and in particular the evolution of highly obscured to Compton-thick ($N_{\rm H} \simgreat \sigma_{\rm T}^{-1} \sim 10^{24}$~\cmsq) sources.  

There is, therefore, strong motivation for extending sensitive X-ray surveys to high energy.   Simple extrapolations of AGN populations detected by \chandra\ and \xmm\ to higher energies based on average spectral properties fail to reproduce the shape and intensity of the CXB at 30~keV \citep[e.g.][]{gc+07}.  This indicates either that spectral models based on the small samples of high-quality 0.1 - 100~keV measurements used in CXB synthesis models fail to capture the true spectral complexity of AGN,   or that an additional highly obscured AGN population is present in the redshift range $0 \simless z \simless 2$.   It is likely that both factors are important at some level.   A higher fraction of reflected emission than typically assumed, which hardens the emission above 10~keV, appear to be present in moderately obscured, $23 < \log (N_{\rm H}/$cm$^{-2}) < 24$, AGN~\citep{rw+11}.  Even at moderate redshifts, reflection fractions are difficult to constrain with data restricted to $< 10$~keV \citep[e.g.][]{dma+14}.    In addition, it is difficult at low redshifts to properly measure high obscuring columns, which can lead to large errors in estimating intrinsic AGN luminosities \citep[e.g.][]{lga+15}.  It is also
challenging to identify Compton-thick objects in the range $0 \simless z \simless 2$ with the limited bandpass of \chandra\ and \xmm\ (at $z\simgreat$2 these missions sample rest frame energies above 20~keV).  Thus both to characterize highly obscured AGN and constrain their evolution at  $z\simless 2$ requires sensitive surveys at energies above 10~keV.

The {\it Nuclear Spectroscopic Telescope Array (NuSTAR)}, the first focusing high-energy X-ray (3 - 79~keV) telescope on orbit \citep{hcc+12}, has been executing a series of extragalactic surveys as part of its core program, with aim of measuring the demographics and properties of obscured AGN. Through contiguous surveys in fields with existing multi-wavelength data combined with dedicated spectroscopic followup of sources serendipitously identified in individual fields, the {\em NuSTAR} extragalactic surveys have improved the sensitivity limits in the hard, 8 -- 24~keV band by two orders of magnitude relative to {\em INTEGRAL} or
{\em Swift}/BAT, and have probed a wide range in redshift, out to $z\simgreat 2$.

In this paper we present the X-ray number counts (log~$N$ - log~$S$) of AGN, including the first sensitive measurements in the 8 -- 24~keV band.  We reach fluxes of $S$(8 -- 24~keV)$ \sim 3 \times 10 ^{-14}$ \fluxu, a factor of 100 deeper than previous measurements in this band.    We compare the number counts to predictions from X-ray background synthesis models, and to extrapolations from {\em Chandra} and {\em XMM-Newton} surveys, and  compare the intensity of resolved sources to that of the CXB.    A companion paper \citep{aab+15} presents direct constraints on the $>10$~keV AGN X-ray luminosity function.  We adopt a flat cosmology with $\Omega_{\Lambda} = 0.7$ and $h = 0.7$, and quote 68.3\% (i.e. 1$\sigma$ equivalent) errors unless otherwise noted.

\section{Observations and Data Reduction}

\begin{figure}
\includegraphics[width=\columnwidth]{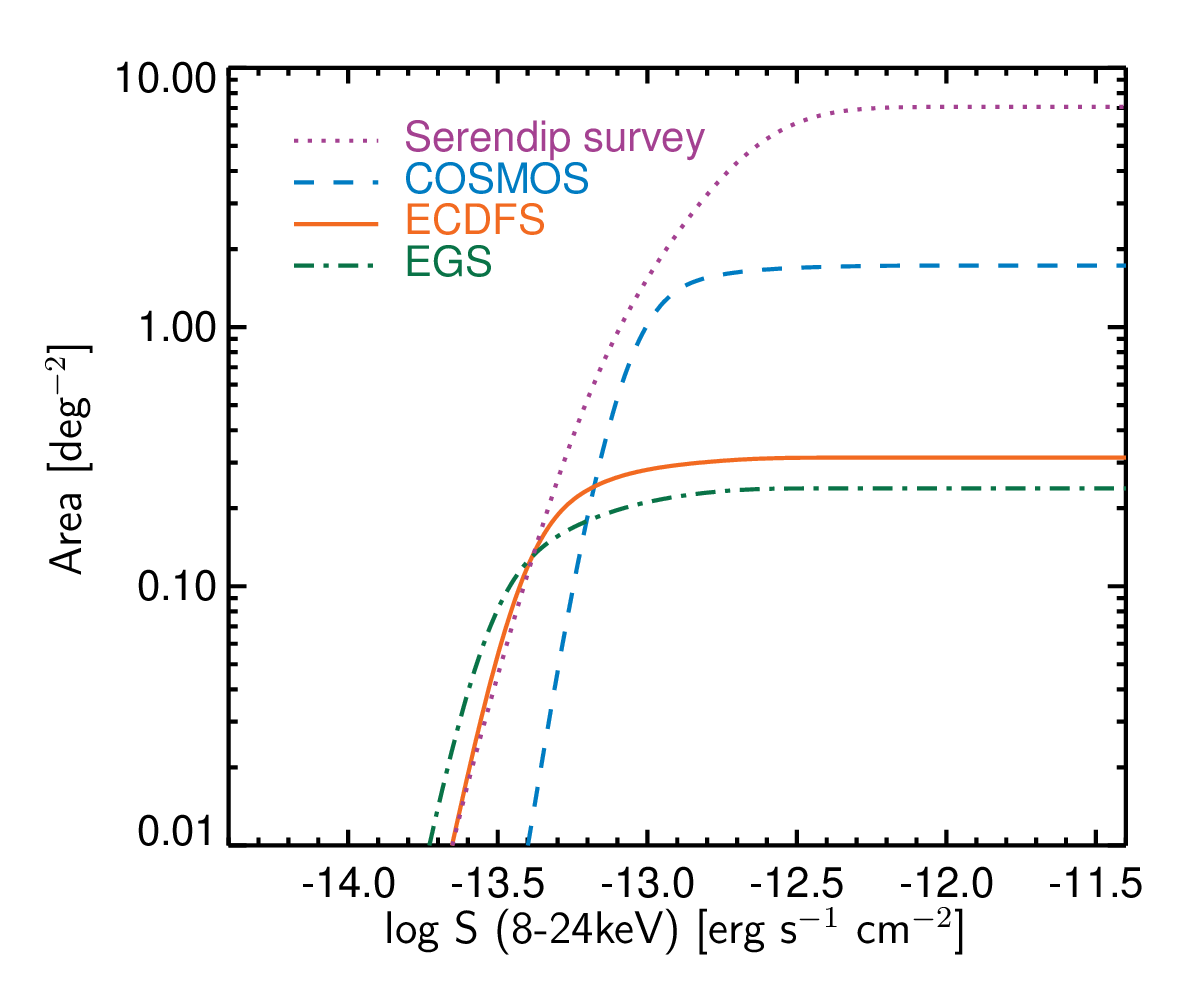}
\caption{Area as a function of depth in the observed 8 -- 24~keV band for the {\em NuSTAR} extragalactic surveys included in this work.}
\label{fig:areavsdepth}
\end{figure}

For the results in this paper we include all components of the \emph{NuSTAR} extragalactic survey program that were completed and analyzed prior to March 2015.
This includes coverage of well-studied contiguous fields as well as identification and followup of sources serendipitously detected in all {\em NuSTAR} fields.  Figure~\ref{fig:areavsdepth} shows the area as a function of depth for the survey components included in this work.
At the shallow end, {\em NuSTAR} surveyed 1.7~deg$^2$ of the Cosmic Evolution Survey field (COSMOS; \citet{sab+07}) to a depth of 
$S$(8 -- 24~keV) = $1.3 \times 10^{-13}$~\fluxu. The catalog and results from this survey are presented in \citet{chp+15} (hereafter C15).    At the deep end, \nustar\ surveys have reached depths of  $S$(8 -- 24~keV) = $2.5 \times 10^{-14}$~\fluxu\ in the Extended Chandra Deep Field South \citep[ECDFS;][]{lba+05} over an area of 0.3~deg$^2$ and in the deep {\em Chandra} region of the Extended Groth Strip \citep{gfh+12,nla+15} over an area of 
0.23 deg$^2$.     The ECDFS source catalog is presented in \citet{mda+15} (hereafter M15) and the EGS catalog will be presented in Aird et al. (in prep).  
The serendipitous survey adopts a similar approach used with the {\em Chandra} SEXSI survey \citep{hem+03}, where fields are searched for point sources not associated with the primary target, and subsequent spectroscopic followup with the Keck, Palomar 200-in, Magellan, and NTT telescopes provides redshifts and AGN classifications.  Preliminary results from this survey were published in \citet{asd+13},  and the 30-month catalog and results will be published in Lansbury et al. (in prep). 

\begin{figure}
\includegraphics[width=\columnwidth]{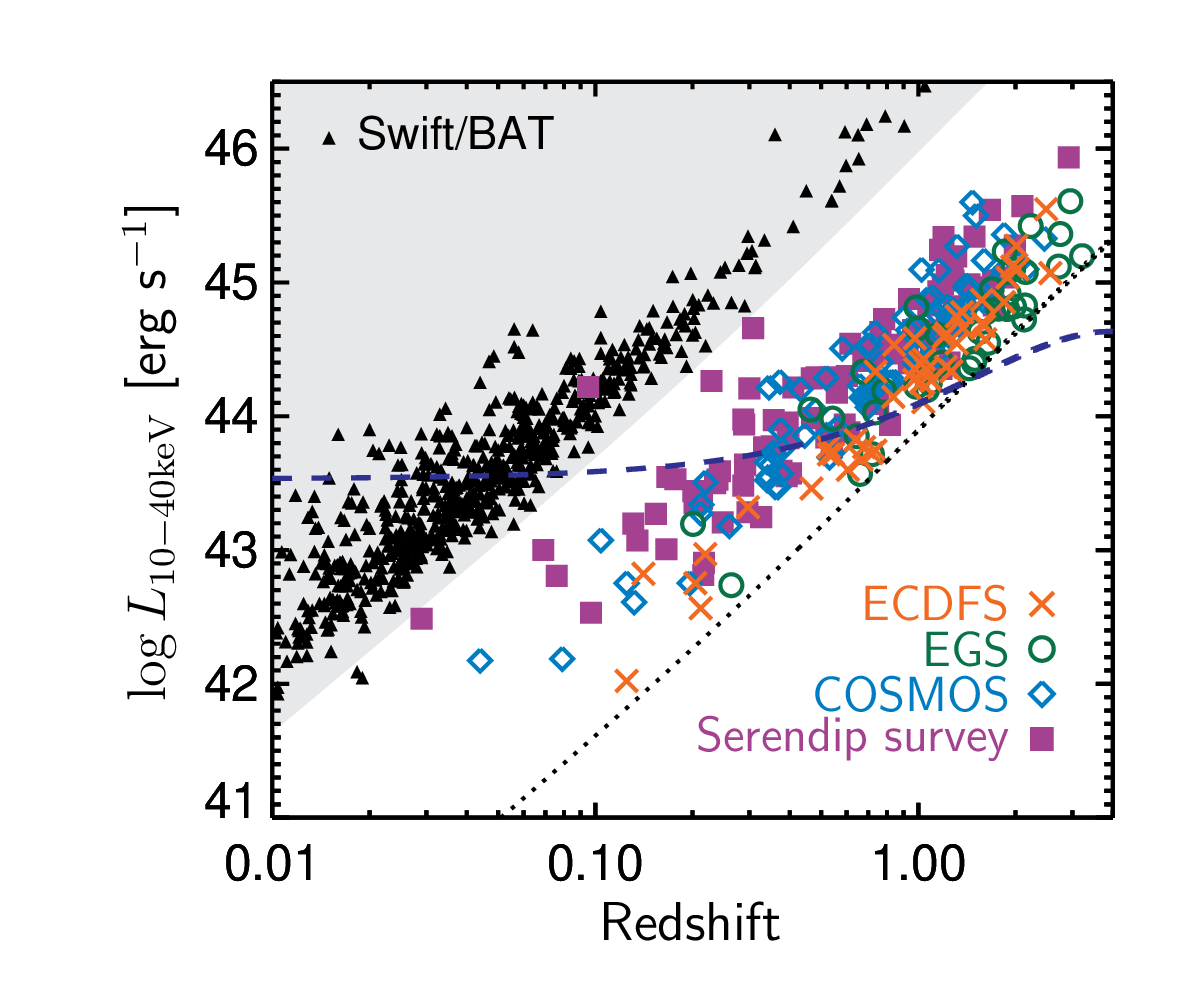}
\caption{Restframe 10 -- 40~keV X-ray luminosity versus redshift for the objects included in this work compared to sources in the \swift/BAT 70-month all-sky survey catalog (black triangles).   The dashed line shows the location of the knee in the luminosity function from \cite{acg+15} as a function of redshift.   The shaded region indicates the region of sensitivity of {\em Swift}/BAT, and the dotted line indicates the threshold for the {\em NuSTAR} surveys. }
\label{fig:logLlogZ}
\end{figure}

The overall sample consists of 382 unique sources detected in the full (3 -- 24~keV), soft (3 -- 8~keV) or hard (8 -- 24~keV) bands, of which 124 are detected in the hard band and 226 are detected in the soft band.
Figure~\ref{fig:logLlogZ} shows the distribution of rest frame 10 -- 40~keV luminosity versus redshift for the sources in our sample.  Luminosities in this plot are derived from the 8 -- 24~keV count rates (if detected in that band) assuming an unabsorbed X-ray spectrum with a photon index $\Gamma = 1.8$ 
folded through the \nustar\ response. 
If the source is not detected in the 8 -- 24~keV band we calculate luminosities using the 3 -- 24~keV or (if not detected at 3 -- 24~keV) the 3 -- 8~keV count rates.
The median redshift for the entire \nustar\ sample is $<z>=0.76$, with a median luminosity of $<\log (L_\mathrm{10-40\;keV}$/erg s$^{-1})> = 44.37$.  For comparison we plot the distribution of AGN in the \swift/BAT 70-month catalog~\citep{btm+13}.  For reference, the dashed line in Figure~\ref{fig:logLlogZ} shows the location of the knee in the luminosity function as a function of redshift from the \textit{Chandra}-based study of \cite{acg+15}, extrapolated to the 10 -- 40~keV band.  Together, these surveys cover a broad range of luminosity and redshift.

\subsection{Contiguous survey fields}

We have adopted uniform source detection and flux extraction methodologies across the COSMOS, ECDFS and EGS survey fields.   For the ECDFS and COSMOS fields we use the source lists and supporting data products (mosaic images, exposure maps and  background maps in the 3 -- 24~keV, 3 -- 8~keV and 8 -- 24~keV energy bands) from M15 and C15, respectively. 
We adopt exactly the same approach for the analysis of the EGS field (Aird et al., in prep.).
We summarize the analysis approach here, and refer the reader to the relevant catalog papers for details.  

For source detection we convolve both the mosaic images and background maps with a 20\arcsec-radius aperture at every pixel, and determine the 
probability, based on Poisson statistics, that the total image counts are produced by a spurious fluctuation of the background.
We generate the background maps based on the NUSKYBGD code \citep[see][for details]{whm+14}.
We then identify groups of pixels in these false probability maps, using SExtractor, where the probability is less than a set threshold.
Different thresholds are used in each field and in each energy band based on the expected number of spurious sources in simulations (see C15, M15 and Aird et al. in prep for specific values used for each field).
We merge detections in multiple bands to produce the final catalogs, which include 61 sources (3 -- 8 keV) and 32 sources (8 -- 24 keV) in COSMOS, 
33 sources (3 -- 8 keV) and 19 sources (8 -- 24 keV) in ECDFS, and 26 sources (3 -- 8 keV) and 13 sources (8 -- 24 keV) in EGS.  The detection thresholds are chosen to ensure the catalogs are 99\% reliable,  Thus, we expect very few spurious sources (see M15, C15). 

Source confusion must be corrected for since blended sources result in mis-estimation of source counts and fluxes.  The \nustar\ PSF is relatively large compared to \chandra\ and \xmm, and so source blending is particularly important to account for when comparing number counts from \nustar\ to these lower-energy, higher-resolution missions.   
The source de-blending procedure is described in detail in M15 and C15.    To test the validity of the procedure we performed Monte Carlo simulations, where source fluxes were drawn from a published number counts distribution,  counts maps were simulated,  and sources were extracted and de-blended.   We then verified that the resulting number counts distribution matches the input.   The details of these simulations are provided in \S4 of C15.

\subsection{Serendipitous survey}

The source-detection procedure for the serendipitous survey is the same as that outlined above, and described in detail in C15 and M15, although with
a slightly different procedure for background determination.  Many of the fields have bright central targets that contaminate a 
portion of the \emph{NuSTAR} field-of-view.   We thus take the original images and convolve them with an annular aperture of inner radius 30\arcsec\
and outer radius 90\arcsec.   We rescale the counts within this annulus  to that of a 20\arcsec\ radius region based on the ratio of the aperture areas and effective exposures.  This procedure produces maps of the local background level at every pixel based on the observed images, 
and will thus include any contribution from the target object. 
We then use these background maps, along with the mosaic images, to generate false probability maps.  
From here the source detection follows that used in the contiguous fields, using a false probability threshold of $<10^{-6}$ across all bands. 
We exclude any detections within 90\arcsec\ of the target position, and
also any areas occupied by large, foreground galaxies or known sources that are associated with the target (but are not at the aimpoint).
We also exclude areas where the effective exposure is $<20$\% of the maximal (on-axis) exposure in a given field.
Any \emph{NuSTAR} fields at Galactic latitude $<20^\circ$ are also excluded from our sample. 
Full details of the serendipitous survey program will be provided in Lansbury et al. (in prep), which will also indicate those serendipitous fields used in
this work.  In total we include 106 serendipitous sources (3 -- 8 keV) and 60 (8 -- 24~keV).

\section{Number count measurements}

\begin{figure*}
\begin{center}
\includegraphics[width=18cm]{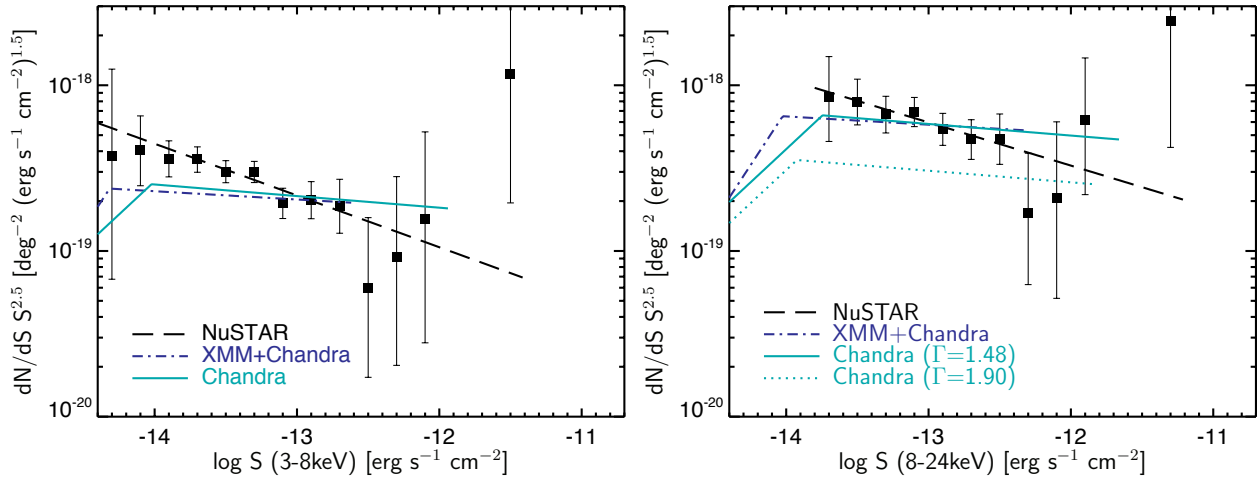}
\caption{{\bf Left panel:} The differential number counts for the observed 3 -- 8~keV band.  The data points (black squares)  show measurements from the combined \nustar\ survey fields, where error bars are 1$\sigma$ equivalent (i.e. 68.3\% confidence level), and the black dashed line shows the best-fit power law.   The solid line shows the best-fit to the number counts as measured by \chandra\  \citep{gnl+08}, and the dot-dashed line shows the best fit from a combined analysis of \xmm\ and \chandra\ data \citep{mwc+08}.     {\bf Right panel:} The differential logN-logS in the 8 -- 24 keV band  (black squares).  The solid blue line and dot-dashed lines show the \chandra\ (4 -- 7~keV) and \xmm\  measurements extrapolated to the 8 -- 24~keV band using a photon power law index of $\Gamma = 1.48$.    The blue dotted line shows the \chandra\ measurements using a steeper ($\Gamma = 1.9$) power law for the spectral extrapolation.}
\label{fig:dnds}
\end{center}
\end{figure*}

To determine the number counts (log~$N$-log~$S$), we adopt a Bayesian approach (see Georgakakis et al. 2008, Lehmer et al. 2012)
that assigns a range of possible fluxes to a given source based on the Poisson distribution, which we then fold through the differential number counts distribution.
We assume the differential number counts are described by a single power-law function;
\begin{equation}
\frac{dN}{dS} =	K \left(\frac{S}{10^{-13}~\rm{erg~s^{-1}~cm^{-2}}}\right)^\beta
\end{equation}
where $K$ is the normalization at $S=10^{-13}$~\fluxu\ and $\beta$ is the slope.
Folding the Poisson likelihood for each individual source through the differential number counts given by Equation 1 accounts for the Eddington bias, allowing for the fact that a detection is more likely to be due to a positive fluctuation from a source of lower flux than vice-versa.
We limit the range of possible source fluxes to a factor 3 below the nominal flux limit\footnote{We note that allowing for a source fluxes a factor 10 or more below the nominal flux limit has a neglible impact on our results.} to prevent the probability distribution from diverging at the faintest fluxes as a result of our assumed power-law function.

We optimize the values of the parameters describing the power-law model for the differential number counts by performing an un-binned maximum likelihood fit for all sources detected in a given band (see Georgakakis et al. 2008).  We also estimate differential source number counts in a number of fixed-width bins in flux using the $N_\mathrm{obs}/N_\mathrm{mdl}$ method of \citet{mhs01}, as expanded on in \citet{anl+10}, to account for flux probability distributions. 
The binned estimate of the differential source number counts is then given by
\begin{equation}
\left[\frac{dN}{dS}\right]_\mathrm{bin} = \left[\frac{dN}{dS}\right]_\mathrm{mdl} \frac{N_\mathrm{obs}}{N_\mathrm{mdl}}
\end{equation}
where $\left[\frac{dN}{dS}\right]_\mathrm{mdl}$ corresponds to the power-law model for the differential number counts evaluated at the center of the bin, $N_\mathrm{mdl}$ is the predicted number of sources in a bin (found by folding the model through the sensitivity curve) and $N_\mathrm{obs}$ is the effective observed number of sources, allowing for the distribution of possible fluxes (thus a single source can make a partial contribution to multiple bins). 
We estimate errors based on Poisson uncertainties in $N_\mathrm{obs}$ as given by \citet{geh86}.

Figure~\ref{fig:dnds} (left) shows the resulting differential source number densities based on the \emph{NuSTAR} samples from the four survey components combined, in the 3--8 keV band.  
We also compare the \nustar\ measurements to the best-fit broken power-law functions determined from \chandra\ surveys by \citet{gnl+08} and a joint analysis of \xmm\ and \chandra\ surveys by \citet{mwc+08}. 
We convert the \chandra\ (4 -- 7~keV)  and \xmm\ (2 -- 10~keV) number counts to match the  3--8~keV \emph{NuSTAR} band assuming a $\Gamma=1.8$ power-law X-ray spectrum.  Because the bands are largely overlapping, the choice of photon spectral index does not significantly
affect the results. 
To ease comparison between the different results we have scaled $dN/dS$ by the Euclidean slope, $S^{-2.5}$.  The \nustar\ measurements constrain the slope well over the range $-14 < \log (S /$\fluxu$) < -12.5$.   The best fit values for the power law parameters in this band 
are: log~$K = 13.84 \pm 0.04$, $\beta = -2.81 \pm 0.08$.   

\begin{figure*}
\begin{center}
\includegraphics[width=18cm]{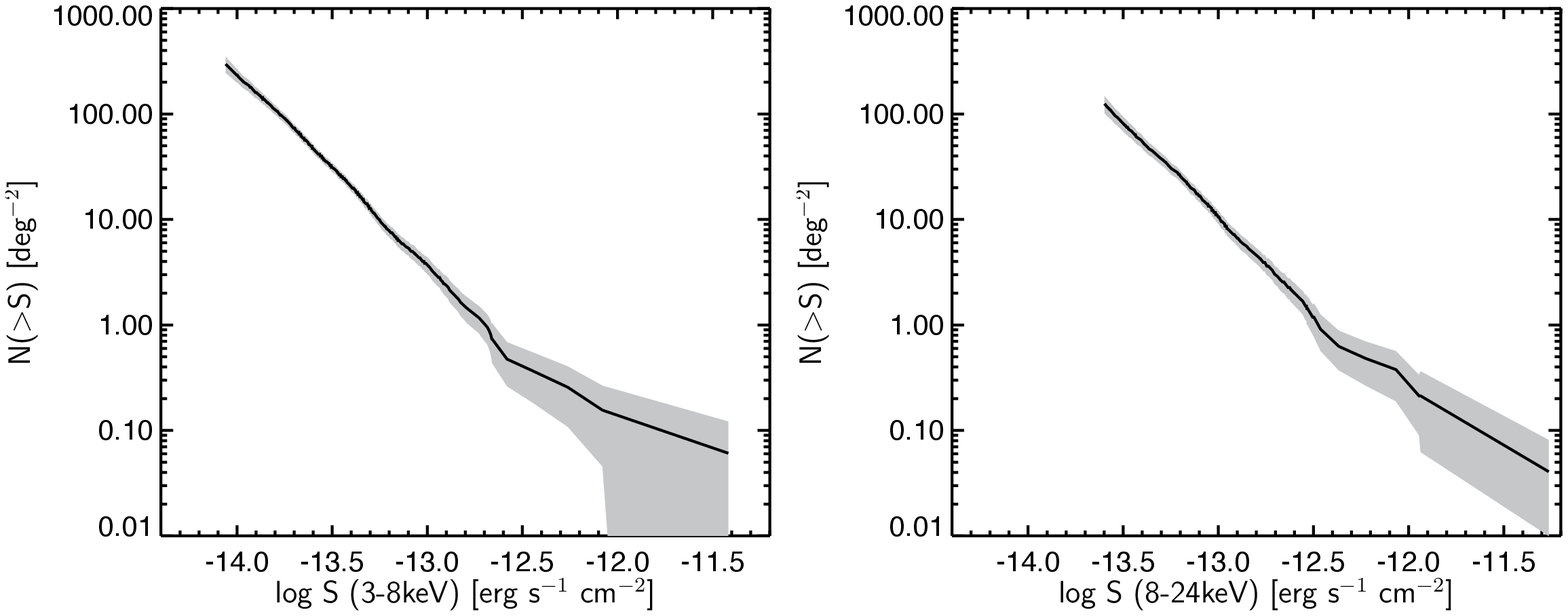}
\caption{Integral number counts for the 3 -- 8 keV (left) and 8 -- 24 (right) observed bands.  The grey shaded region shows the 68.3\% confidence region on the integrated
number counts based on the Poisson error in the number of sources weighted by the survey area as a function of flux.  The blue line on the left plot shows a comparison to \chandra\ counts extrapolated from 4 -- 7~keV.}
\label{fig:ngts}
\end{center}
\end{figure*}

The \nustar\ results generally agree with \chandra\ and \xmm,
although the slope of the number counts with flux is somewhat steeper.  Below $S \sim 10^{-14}$~\fluxu\ the \nustar\ measurements are poorly constrained, so that the location of the break in the number counts seen by \chandra\ and \xmm\ cannot be independently confirmed.    The difference in slope of the number counts with flux between \nustar\ and the soft X-ray telescopes is due the different instrument responses and the corresponding uncertainties in converting count rates to fluxes
based on simple spectral models.     To verify this we folded the population synthesis model of \citet{acg+15} through the \nustar\ response function to predict the
observed count-rate distribution.   We convert the count rates to fluxes, applying our standard count rate to flux conversion factor.   We repeated this exercise adopting
the \chandra\ response function, applying the count rate to flux conversion factor from Georgakakis et al. (2008).  For moderately to heavily absorbed
sources ($N_\mathrm{H} \sim 10^{23}$ cm$^{-2}$), the expected count rates are more strongly suppressed using the Chandra response (as the
\nustar\ instrument response is more strongly weighted to higher
energies). The intrinsic fluxes are therefore underestimated for such sources with \chandra\ when a single conversion factor is assumed. We have
verified that this effect leads to an slope difference in the expected log $N$ -- log $S$ at $f_\mathrm{3-8keV}\sim10^{-14}-10^{-13}$ erg s$^{-1}$cm$^{-2}$ that matches the discrepancy seen between the NuSTAR measurements and the  \chandra\ and \xmm\ measurements shown in Figure 3 (left).
Figure~\ref{fig:ngts} (left) shows the integral number counts in the 3 -- 8~keV band.

 Figure~\ref{fig:dnds} (right) shows the differential source number counts in the 8 -- 24~keV band, along with the best fit power law, parametrized by
 log~$K = 14.3 \pm 0.04$, and $\beta = -2.76 \pm 0.10$.    
 We also plot extrapolations of the \cite{mwc+08} \chandra + \xmm\ counts and the \cite{gnl+08} \chandra\ counts to the harder, largely non-overlapping
 8 -- 24~keV band (dotted and solid lines in Figure~\ref{fig:dnds}).
The dotted line shows the extrapolation assuming $\Gamma = 1.8$, which is typical of unabsorbed AGNs, and systematically underpredicts the \nustar\ measurements.
Using a maximum likelihood analysis, we find that a spectral photon index of $\Gamma = 1.48$ (solid line) provides the best match between the extrapolation of the \citet{gnl+08} model and the \nustar\ data,
indicating that a substantial population of absorbed and/or hard-spectrum sources is required to reproduce the \nustar\ measurements.
  
\section{Comparison with the \swift/BAT Local AGN Sample}

In combination, \nustar\ and \swift/BAT sample the hard X-ray AGN population over a wide range in flux and redshift.    \swift/BAT has measured the number
counts at fluxes $S\simgreat 3 \times 10^{-12}$~\fluxu\ in the 15 -- 55~keV band for the local ($z \simless 0.1$) AGN sample~\citep{aa+12}.
Although there is a gap in the flux range probed by BAT and the $S \simless 3 \times 10^{-13}$~\fluxu\ population probed by \nustar,
extrapolating the BAT differential number counts to fainter fluxes indicates whether there is any evolution between the low-redshift BAT AGN population
and the higher-redshift \nustar\ sources.

\begin{figure}
\begin{center}
\includegraphics[width=\columnwidth]{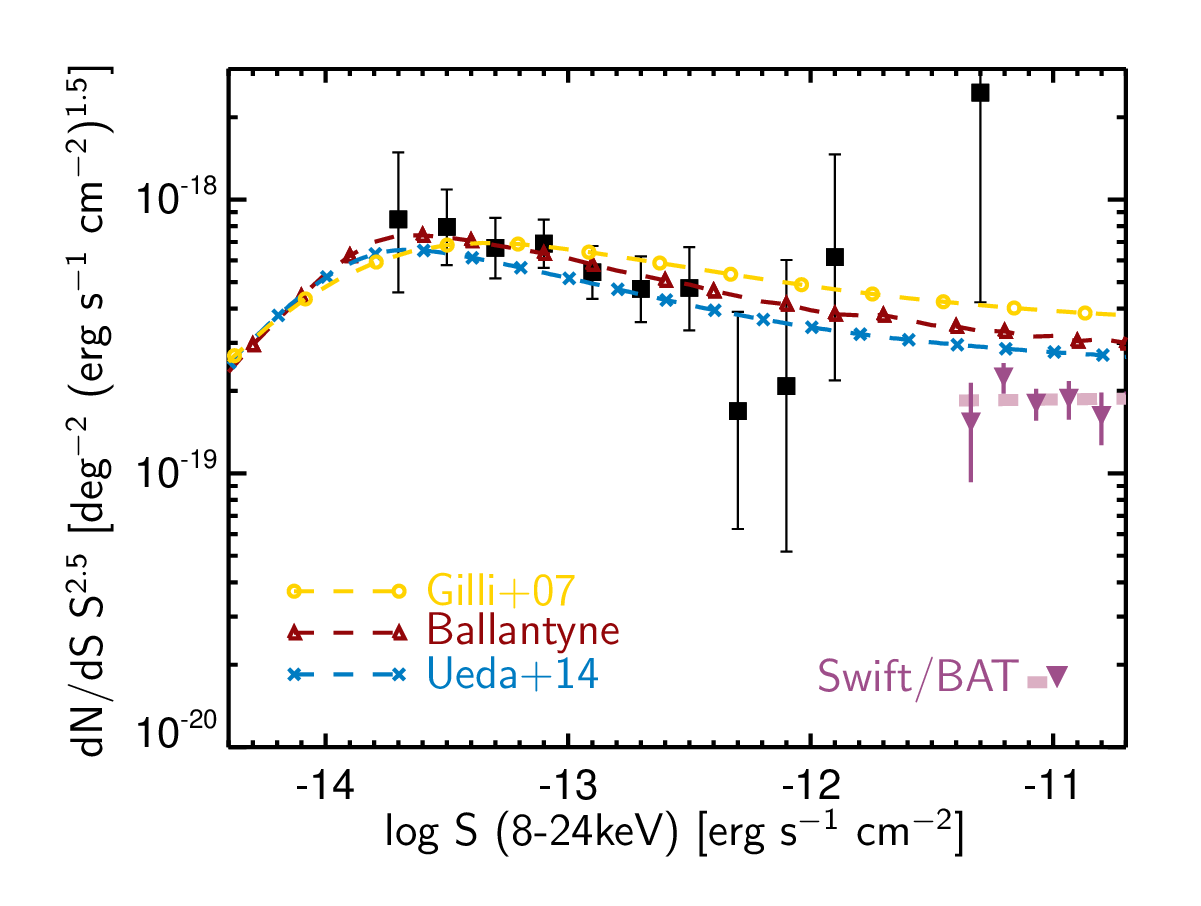}
\caption{The \nustar\ 8 -- 24~keV number counts compared to \swift/BAT (bold dashed line) (Ajello et al. 2012).    The hatched region indicates the uncertainty in the overall fit from Ajello et al. (2012).  We convert the BAT 15-55~keV band to the 8 -- 24~keV band using a power law with photon index
$\Gamma = 1.7$ (the best-fit to the average BAT AGN spectrum in the 15 -- 55~keV band).   The dashed line with triangles shows predictions from the CXB synthesis model of Ballantyne~2011 updated to use the Ueda~2014 luminosity function.  The dashed line with crosses shows the population synthesis model from \cite{uah+14}.  The dashed line with circles shows predictions from the CXB model from \citet{gc+07}.  The plotted errors are 1-$\sigma$.}
\label{fig:bat}
\end{center}
\end{figure}

Figure~\ref{fig:bat} shows the \nustar\ 8 -- 24~keV number counts together with the BAT measurements from \citet{aa+12}, which are well described by a single power-law $dN/dS$ with slope $\beta\approx 2.5$  (bold dashed line).   We have converted the BAT 15 - 55~keV fluxes  to the 8 -- 24~keV band assuming a photon index of $\Gamma = 1.7$,  the value that provides the best fit to the average BAT AGN spectral model from \citet{bag+11}.  The hatched region indicates the measurement uncertainties from \citet{aa+12}, showing that the formal error is small, about the width of the dashed line. 
It is clear that the extrapolation of the BAT counts to lower fluxes where the counts are well-constrained by \nustar\ ($f \sim 10^{-13.5}$\fluxu\ significantly under-predicts the \nustar\ measurements.  This disagreement is not surprising, since there is strong evolution in the AGN population between the local ($z\sim0.1$) BAT sample and the higher-redshifts probed by \nustar.

\section{Comparison with X-ray Background Synthesis Models}

The existence of a population of obscured and Compton thick objects beyond those directly resolved below 10~keV has long been postulated by
CXB synthesis models.  These models reproduce the hard spectrum of the CXB using measured AGN X-ray luminosity functions (XLFs), which are well constrained (except in the very local universe) only below 10~keV, together with models for the broadband  ($\sim$1 -- 1000~keV) AGN spectra 
and estimates of the Compton-thick AGN fraction and its evolution (e.g. \citet{gc+07}, \citet{tuv09}, \citet{uah+14}). The XLF measurements, assumptions about AGN spectral evolution, and Compton-thick fractions differ significantly among models.

Figure~\ref{fig:bat} compares the \nustar\ counts to three different CXB synthesis models: the model from \citet{gc+07} (dashed line with circles), the model of \citet{uah+14}, and an an updated version of the Ballantyne (2011) model.   The details of the \citet{gc+07}  and  \citet{uah+14} models are documented in the relevant publications.   The updated Ballantyne model differs from Ballantyne (2011) in that it uses the new Ueda et al. (2014) luminosity function, a better spectral model from Ballantyne (2014), the Burlon et al. (2011) N$_{\rm H}$ distribution, and the redshift evolution and obscured fraction from Ueda et al. (2014).    Other aspects, including the normalization of the Compton-thick fraction are the same as in Ballantyne (2011).  We include this model because it fits the CXB spectrum even after including the effects of blazers.  

All three models are in good agreement with the measured \nustar\ counts.  However, all of the models lie significantly above the \swift/BAT measurements 
at $S \simgreat 10^{-11.5}$\fluxu.   The discrepancy between the BAT measurements and the models cannot be accounted for by
uncertainties in the spectral shape; a spectral index of $\Gamma = 2.1$ when converting to the 8 - 24~keV band is required to make the \swift/BAT 15 -- 55~keV measurements agree with the model predictions.  This is significantly softer than the average measured spectral shape, even for unobscured AGN, and can 
thus be ruled out.   Therefore, this discrepancy appears to be due to evolution in the hard XLF, absorption distribution, or spectral properties of AGN between the very local \swift/BAT sample and the more distant ($z \sim 0.5 - 1$) \nustar\ sample that is not fully accounted for in these population synthesis models (see
also Aird et al. 2015b).

\section{Summary and Conclusions}

\begin{figure}
\begin{center}
\includegraphics[width=\columnwidth]{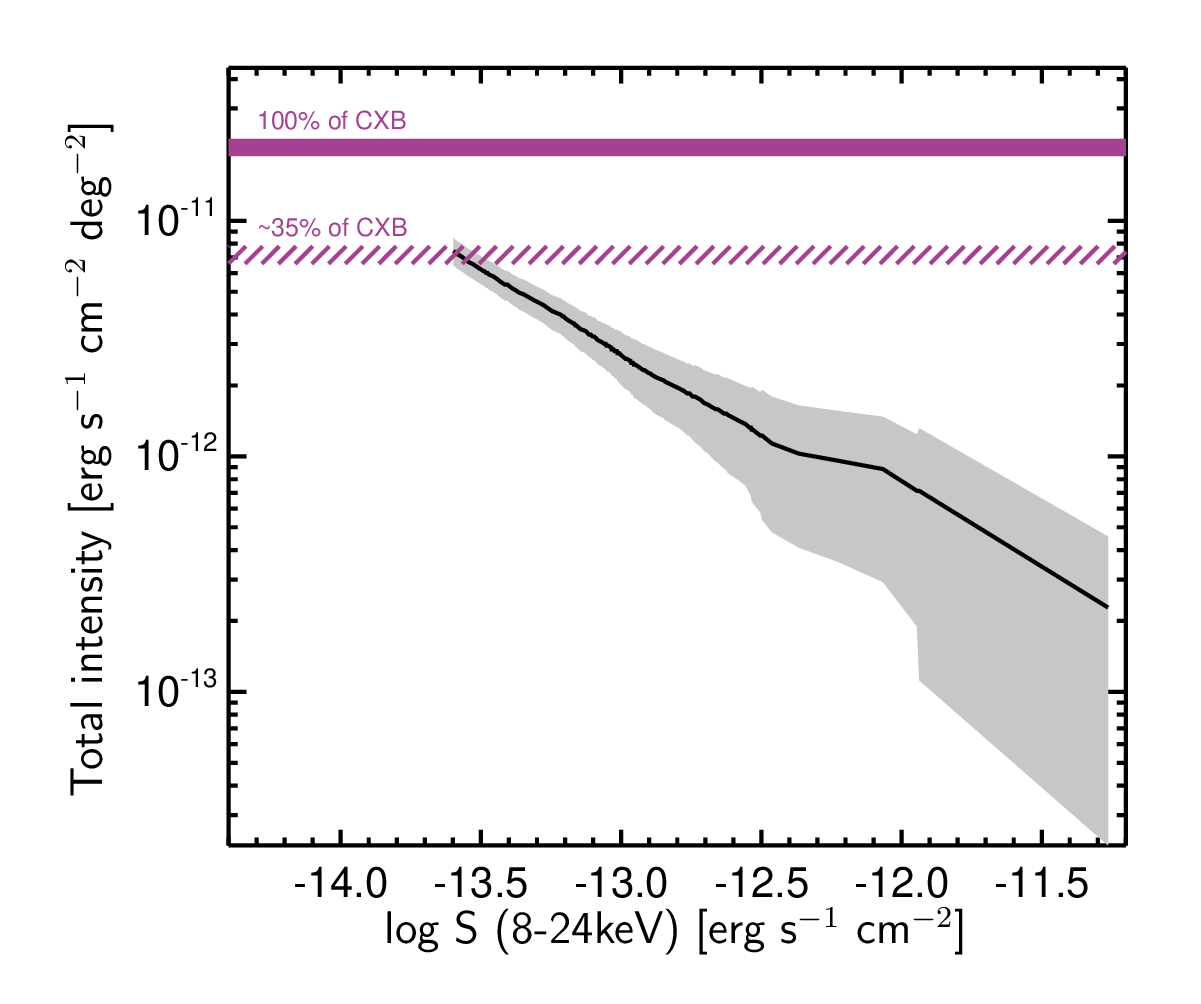}
\caption{Total intensity as a function of 8 - 24~keV flux of the resolved sources in the sample included in this work.   The horizontal lines indicate 35\% (hatched line) and 100\% (solid line) of the cosmic X-ray background (CXB).   The width of the lines indicates the range of normalizations determined by different instruments (see Table~\ref{tab:cmb_comp} for references).   
At the faint end of the surveys included here \nustar\ is resolving $\sim$35\% of the CXB.}
\label{fig:intcxb}
\end{center}
\end{figure}

We have presented measurements of the number counts of AGN with \nustar\  in two bands, from 3 -- 8~keV and 8 -- 24~keV.   The data span a broad range in flux, with good constraints covering $10^{-14} \simless $ S (8 - 24~keV; \fluxu) $\simless 10^{-13}$.  The 3 -- 8~keV differential source number densities are in agreement with
measurements from \chandra\ and \xmm, although the slope measured by \nustar\ is somewhat steeper.  The slope difference results from the 
fact that the \nustar\ effective area curve is weighted to significantly higher energies compared to \xmm\ and \chandra, which results in the observed 
discrepancy when using a simple counts to flux conversion factor.

\begin{deluxetable*}{lccc} 
\tabletypesize{\footnotesize} 
\tablecolumns{4} 
\tablewidth{0pt} 
\tablecaption{ X-ray Background Measurements \label{tab:cmb_comp}} 
\tablehead{\colhead{Instrument} & \colhead{$I_{CXB}$ (20 - 50 keV)} & \colhead{$I_{CXB}$ (8 - 24 keV)} & \colhead{\% resolved by {\em NuSTAR}}}
\startdata
{\em HEAO-1} A2 + A4 &  $6.06 \pm 0.06 $ & $6.33 \pm 0.07 $  &   39   \\
{\em HEAO-1} A2  & $5.60 \pm 0.30 $  &$6.27 \pm 0.33 $ &  39 \\
{\em BeppoSAX}  &  $5.89 \pm 0.19 $  & $6.48 \pm 0.21 $ &  38  \\
{\em INTEGRAL}  & $\sim 6.66 $  &  $7.33 $ &  33  \\
{\em BAT} & $6.50 \pm 0.15 $  &  $7.16 \pm 0.17 $ &  34  \\
\enddata
\tablecomments{CXB intensity ($I_{CXB}$) is given in units of $10^{-8}$ergs~cm$^{-2}$~s$^{-1}$~sr$^{-1}$. Intensities in the 20 -- 50~keV band are taken from \citet{gmp+99} for {\em HEAO-1} A2 + A4, \citet{mbh+80} for {\em HEAO-1} A2,
\citet{fol+07} for {\em BeppoSAX}, \citet{csr+07} for {\em INTEGRAL}, and \citet{ags+08} for {\em BAT}. }
\end{deluxetable*}

In the 8 -- 24~keV band we present the first direct measurement of the AGN number counts that includes data above $\sim$10~keV and reaches down to flux levels $\sim 3 \times 10^{-14}$ \fluxu.   In order to match the
\nustar\ number counts, the flux measurements from \chandra\ and \xmm\ must be extrapolated to higher energy using a spectrum with photon index $\Gamma = 1.48$.  This photon index is significantly harder than the $\Gamma = 1.7 - 1.9$ that characterizes the unobscured AGN population, and is also significantly harder than the average spectral index that characterizes the  \swift/BAT AGN sample.   

The \nustar\ number counts are in good agreement with predictions from population synthesis models that explain the hard spectrum of the CXB using different assumptions
about obscuration, the Compton thick sample, and the spectra shape of AGN in the hard X-ray band (see Figure~\ref{fig:bat}).
This directly confirms the existence of a population of AGN with harder spectra than those typically measured below 10~keV.   The spectral hardness could be due
either to increased reflection or near Compton-thick absorption.   The updated Ballantyne model, for example, uses an AGN spectral model with a strong Compton reflection component with a relative normalization component of $R = 1.7$.  Our
measurements of the rest-frame 10 -- 40~keV XLF~\cite{aab+15} also indicate that a significant population of AGN with hard X-ray spectra is required to reconcile our
\nustar\ data with prior, lower energy XLF measurements.
To what extent this results from obscuration versus higher levels of reflection will be determined by spectral analysis of \nustar\ sources in the survey fields (Del Moro et al. in prep, Zappacosta et al. in prep), and from spectral modeling of high quality data from local AGN samples (Balokovi\'{c} et al. in prep).

The \nustar\ 8 -- 24~keV number counts lie significantly above a direct extrapolation with flux of the number counts at brighter fluxes, sampled by the \swift/BAT survey in the 15 -- 55~keV band (Figure~\ref{fig:bat}).   This discrepancy is not surprising given the known evolution of the AGN population between the low redshift \swift/BAT
AGN and the higher redshift \nustar\ sample   It is interesting that the BAT data are  in tension with CXB synthesis models (see Figure~\ref{fig:bat}), which are in good agreement with the \nustar\ measurements.    The most natural explanation for the difference is an evolution in the hard XLF, absorption distribution, or spectral properties of AGN between the very local objects seen by BAT and the more distant ($z\sim 0.5 - 1$) \nustar\ sample that is not accounted for in the current population synthesis models.

Table~\ref{tab:cmb_comp} provides CXB fluxes measured by hard X-ray instruments in the  20 - 50~keV and 8 -- 24~keV bands.  To convert the intensities to the 8 -- 24~keV band
for those instruments that do not cover this energy range ({\em BeppoSAX, INTEGRAL, Swift}/BAT) we used 
Equation~5 of \citet{ags+08}, which parametrizes the CXB spectrum between 2~keV and 2~MeV based on a fit to available data.    
The \nustar\ extragalactic surveys have reached depths of $S$(8 -- 24~keV)$\approx 3 \times 10^{-14}$ \fluxu.   Comparing to the total integrated flux of the CXB as measured by
collimated and coded aperture instruments shown in Table~\ref{tab:cmb_comp},  this corresponds to a resolved fraction of 33 - 40\% in the 8 -- 24~keV band (Figure~\ref{fig:intcxb}), with an additional
statistical uncertainty of 5\%.    Even for the highest measured CXB flux from \citet{csr+07} \nustar\
still resovles 33\%,  which  is a significant advance compared to the 1 - 2\% resolved to-date by coded mask instruments above 10~keV \citep{vm+13,kr+11}.

The resolved fraction of the CXB we measure is in good agreement with pre-launch predictions based on CXB synthesis models \cite{bdm+11}.  At the current depth the \nustar\ surveys do not probe the break in the number counts distribution, expected, based on extrapolations from
\chandra\ and \xmm, to occur at $S$(8 -- 24~keV)$\sim$10$^{-14}$ \fluxu.   
Reaching these depths will be challenging, as the deep fields are currently
background dominated, such that sensitivity improves only as the square root of observing time.    However, additional exposure in the ECDFS is planned, along with expansion of
the deep surveys to cover the CANDELS/UDS field~\citep{gkf+11}, and
continuation of the serendipitous survey, which will better constrain the slope of the number counts distribution above the break and improve spectral constraints on the resolved AGN population.

\acknowledgments
This work was supported under NASA Contract No. NNG08FD60C, and
made use of data from the {\it NuSTAR} mission, a project led by
the California Institute of Technology, managed by the Jet Propulsion
Laboratory, and funded by the National Aeronautics and Space
Administration. We thank the {\it NuSTAR} Operations, Software and
Calibration teams for support with the execution and analysis of
these observations.  This research has made use of the {\it NuSTAR}
Data Analysis Software (NuSTARDAS) jointly developed by the ASI
Science Data Center (ASDC, Italy) and the California Institute of
Technology (USA).  
JA acknowledges support from ERC Advanced Grant FEEDBACK at the University of Cambridge and 
a COFUND Junior Research Fellowship from the Institute of Advanced Study, Durham University.

\bibliographystyle{jwapjbib}


%

\end{document}